\pdfoutput=1
\documentclass[a4paper,twocolumn,english,amsmath,amssymb,aps,prl,reprint,floatfix,superscriptaddress,showpacs,booktabs]{revtex4}
\usepackage[T1]{fontenc}
\usepackage[latin9]{inputenc}
\usepackage{color}
\usepackage{babel}
\usepackage{textcomp}
\usepackage{amsmath}
\usepackage{amssymb}
\usepackage{graphicx}
\usepackage{subscript}
\usepackage[unicode=true,
 bookmarks=true,bookmarksnumbered=false,bookmarksopen=false,
 breaklinks=true,pdfborder={0 0 1},backref=false,colorlinks=false]
 {hyperref}
\hypersetup{
 pdfauthor={A. van Roekeghem et al}}

\makeatletter

\pdfpageheight\paperheight
\pdfpagewidth\paperwidth

\providecommand{\tabularnewline}{\\}

\@ifundefined{textcolor}{}
{%
 \definecolor{BLACK}{gray}{0}
 \definecolor{WHITE}{gray}{1}
 \definecolor{RED}{rgb}{1,0,0}
 \definecolor{GREEN}{rgb}{0,1,0}
 \definecolor{BLUE}{rgb}{0,0,1}
 \definecolor{CYAN}{cmyk}{1,0,0,0}
 \definecolor{MAGENTA}{cmyk}{0,1,0,0}
 \definecolor{YELLOW}{cmyk}{0,0,1,0}
}

\makeatother

\begin{document}

\title{Tetragonal and collapsed-tetragonal phases of CaFe\textsubscript{2}As\textsubscript{2}
-- a view from angle-resolved photoemission and dynamical mean field
theory}

\author{Ambroise van Roekeghem}

\email{vanroeke@cpht.polytechnique.fr}

\affiliation{Beijing National Laboratory for Condensed Matter Physics, and Institute
of Physics, Chinese Academy of Sciences, Beijing 100190, China}

\affiliation{Centre de Physique Th\'{e}orique, Ecole Polytechnique, CNRS UMR
7644, 91128 Palaiseau, France}

\author{Pierre Richard}

\affiliation{Beijing National Laboratory for Condensed Matter Physics, and Institute
of Physics, Chinese Academy of Sciences, Beijing 100190, China}

\affiliation{Collaborative Innovation Center of Quantum Matter, Beijing, China}

\author{Xun Shi}

\affiliation{Beijing National Laboratory for Condensed Matter Physics, and Institute
of Physics, Chinese Academy of Sciences, Beijing 100190, China}

\affiliation{Swiss Light Source, Paul Scherrer Insitut, CH-5232 Villigen PSI,
Switzerland}

\author{Shangfei Wu}

\affiliation{Beijing National Laboratory for Condensed Matter Physics, and Institute
of Physics, Chinese Academy of Sciences, Beijing 100190, China}

\author{Lingkun Zeng}

\affiliation{Beijing National Laboratory for Condensed Matter Physics, and Institute
of Physics, Chinese Academy of Sciences, Beijing 100190, China}

\author{Bayrammurad Saparov}

\affiliation{Materials Science and Technology Division, Oak Ridge National Laboratory,
Oak Ridge, Tennessee 37831-6114, USA}

\author{Yoshiyuki Ohtsubo}

\affiliation{Synchrotron SOLEIL, Saint-Aubin-BP 48, F-91192 Gif sur Yvette, France}

\author{Tian Qian}

\affiliation{Beijing National Laboratory for Condensed Matter Physics, and Institute
of Physics, Chinese Academy of Sciences, Beijing 100190, China}

\author{Athena S. Sefat}

\affiliation{Materials Science and Technology Division, Oak Ridge National Laboratory,
Oak Ridge, Tennessee 37831-6114, USA}

\author{Silke Biermann}

\affiliation{Centre de Physique Th\'{e}orique, Ecole Polytechnique, CNRS UMR
7644, 91128 Palaiseau, France}

\affiliation{Coll\`{e}ge de France, 11 place Marcelin Berthelot, 75005 Paris,
France}

\affiliation{European Theoretical Synchrotron Facility, Europe}

\author{Hong Ding}

\affiliation{Beijing National Laboratory for Condensed Matter Physics, and Institute
of Physics, Chinese Academy of Sciences, Beijing 100190, China}

\affiliation{Collaborative Innovation Center of Quantum Matter, Beijing, China}
\begin{abstract}
We present a study of the tetragonal to collapsed-tetragonal transition
of CaFe\textsubscript{2}As\textsubscript{2 } using angle-resolved
photoemission experiments and dynamical mean field theory-based electronic
structure calculations. We observe that the collapsed-tetragonal phase
exhibits reduced correlations and a higher coherence temperature due
to the stronger Fe-As hybridization. Furthermore, a comparison of
measured photoemission spectra and theoretical spectral functions
shows that momentum-dependent corrections to the density functional
band structure are essential for the description of low-energy quasiparticle
dispersions. We introduce those using the recently proposed combined
``Screened Exchange + Dynamical Mean Field Theory'' scheme. 
\end{abstract}
\maketitle

\section{Introduction}

CaFe\textsubscript{2}As\textsubscript{2 } at ambient pressure and
temperature is in a paramagnetic tetragonal phase. When temperature
is lowered under 170 K it develops a collinear antiferromagnetic order
and becomes orthorhombic \cite{Ronning-CaFe2As2,Ni-CaFe2As2,Diallo-CaFe2As2-spin}.
Under pressure, this orthorhombic phase can be suppressed and replaced
by a non-magnetic collapsed-tetragonal phase in which the distance
between two FeAs layers is strongly reduced due to the formation of
covalent bonds between As atoms from two different layers. Superconductivity
can also develop from this collapsed phase \cite{Torikachvili-CaFe2As2-SC,Kreyssig-CaFe2As2-collapsed}.
Recently, it has been found that a quench of the annealing phase during
crystal synthesis can produce samples presenting similar properties
as CaFe\textsubscript{2}As\textsubscript{2} under pressure \cite{Ran-CaFe2As2-collapsed,Saparov-CaFe2As2-collapsed}.
At ambient temperature and pressure, they are in the tetragonal phase,
and when temperature is lowered there is a transition into a collapsed-tetragonal
phase, around 90 K in our samples \cite{Saparov-CaFe2As2-collapsed}.
There are also other ways to induce a collapse transition at ambient
pressure, such as isovalent substitution of As by P \cite{Coldea-CaFe2P2},
electron-doping by Rh at the Fe site \cite{Danura-Ca(FeRh)2As2} or
electron doping by rare-earth on the Ca site \cite{Saha-rare-earth-CaFe2As2}.
However, while CaFe\textsubscript{2}As\textsubscript{2} in the collapsed-tetragonal
phase can become superconductor under pressure 
\footnote{Superconductivity can also occur in a non-collapsed phase, e.g. in
Ca\textsubscript{1-x}La\textsubscript{x}Fe\textsubscript{2}As\textsubscript{2}
\cite{Saha-rare-earth-CaFe2As2} or for low-doping values in CaFe\textsubscript{2}As\textsubscript{2-x}P\textsubscript{x}
\cite{Kasahara-Fermi-liquid-CaFe2As2}.
} or with rare-earth doping \cite{Torikachvili-CaFe2As2-SC,Saha-rare-earth-CaFe2As2},
it is not the case in these quenched crystals.

During the collapse, the $c$ axis of the unit cell is strongly reduced
by about 10\%, while the $a$ axis is enlarged by about 2\%. This
modification of the crystal structure is at the origin of a reorganization
of the Fermi surface and electronic structure of the compound \cite{Coldea-CaFe2P2,Danura-Ca(FeRh)2As2,Tsubota-Ca(FeRh)2As2,Dakha-CaFe2As2,Gofryk-CaFe2As2},
which has been studied within DFT \cite{Yildirim-spin-As,Tomic-CaFe2As2,Coldea-CaFe2P2,BaCo2As2-Dakha}
and very recently within combined density functional dynamical mean
field theory (``DFT+DMFT'') \cite{Mandal-CaFe2As2,Diehl-CaFe2As2}.
In particular, it was found that the electronic correlations are reduced
in the collapsed phase. Interestingly, the resistivity at the transition
changes its low-energy behavior from $\rho\propto T$ or $\rho\propto T^{1.5}$
in the tetragonal phase to $\rho\propto T^{2}$ -- as in a good Fermi
liquid -- in the collapsed-tetragonal phase \cite{Kasahara-Fermi-liquid-CaFe2As2,Danura-Ca(FeRh)2As2,Saparov-CaFe2As2-collapsed}.
In rare-earth electron-doped Ca\textsubscript{1-x}RE\textsubscript{x}Fe\textsubscript{2}As\textsubscript{2 },
this Fermi-liquid like resistivity is also observed at low temperature,
independently of the stable phase -- collapsed-tetragonal or n\textcolor{black}{on-collapsed-tetragonal
as in Ca\textsubscript{1-x}La\textsubscript{x}Fe\textsubscript{2}As\textsubscript{2 }\cite{Saha-rare-earth-CaFe2As2}.}
Recent Nuclear Magnetic Resonance data further indicate a suppression
of antiferromagnetic spin fluctuations in the collapsed-tetragonal
phase \cite{Furukawa-CaFe2As2}. Motivated by these intriguing results,
we have performed DFT+DMFT calculations and ARPES experiments on the
tetragonal and collapsed-tetragonal phase of CaFe\textsubscript{2}As\textsubscript{2}.

\section{The collapse transition as seen by ARPES}

We have performed angle-resolved photoemission measurements on samples
grown by the self-flux method that are quenched from 960\textdegree{}C
(corresponding to as-grown ``p1'' samples in \cite{Saparov-CaFe2As2-collapsed}).
Experiments were conducted at the CASSIOPEE beamline of SOLEIL synchrotron
(France) and at the Institute of Physics, Chinese Academy of Sciences
(China). Both systems are equipped with VG-Scienta R4000 electron
analyzers. All samples were cleaved \textit{in situ} at temperatures
higher than 200 K and measured in a working vacuum between $5\times10^{-10}$
and $1\times10^{-9}$ torr at SOLEIL, and better than $5\times10^{-11}$
torr at the Institute of Physics. The photon energy was varied from
20 to 80 eV in synchrotron while we used the He I$\alpha$ line of
an helium discharge lamp in the lab (21.218 eV). The angular resolution
was better than 0.5\textdegree{} and the energy resolution better
than 10 meV.

Samples were measured at 200 K, 100 K, 80 K and 30 K. The tetragonal
to collapsed-tetragonal transition is shown to occur around 90 K by
our magnetic susceptibility measurements, with a hysteresis smaller
than 5 K, in agreement with \cite{Saparov-CaFe2As2-collapsed}. Although
we have performed the temperature-dependent measurements at the Institute
of Physics, the Fermi surfaces obtained in SOLEIL are similar to those
obtained in our laboratory, indicating that the measured samples are
in the same phase.

\begin{figure}
\begin{centering}
\includegraphics[width=8.5cm]{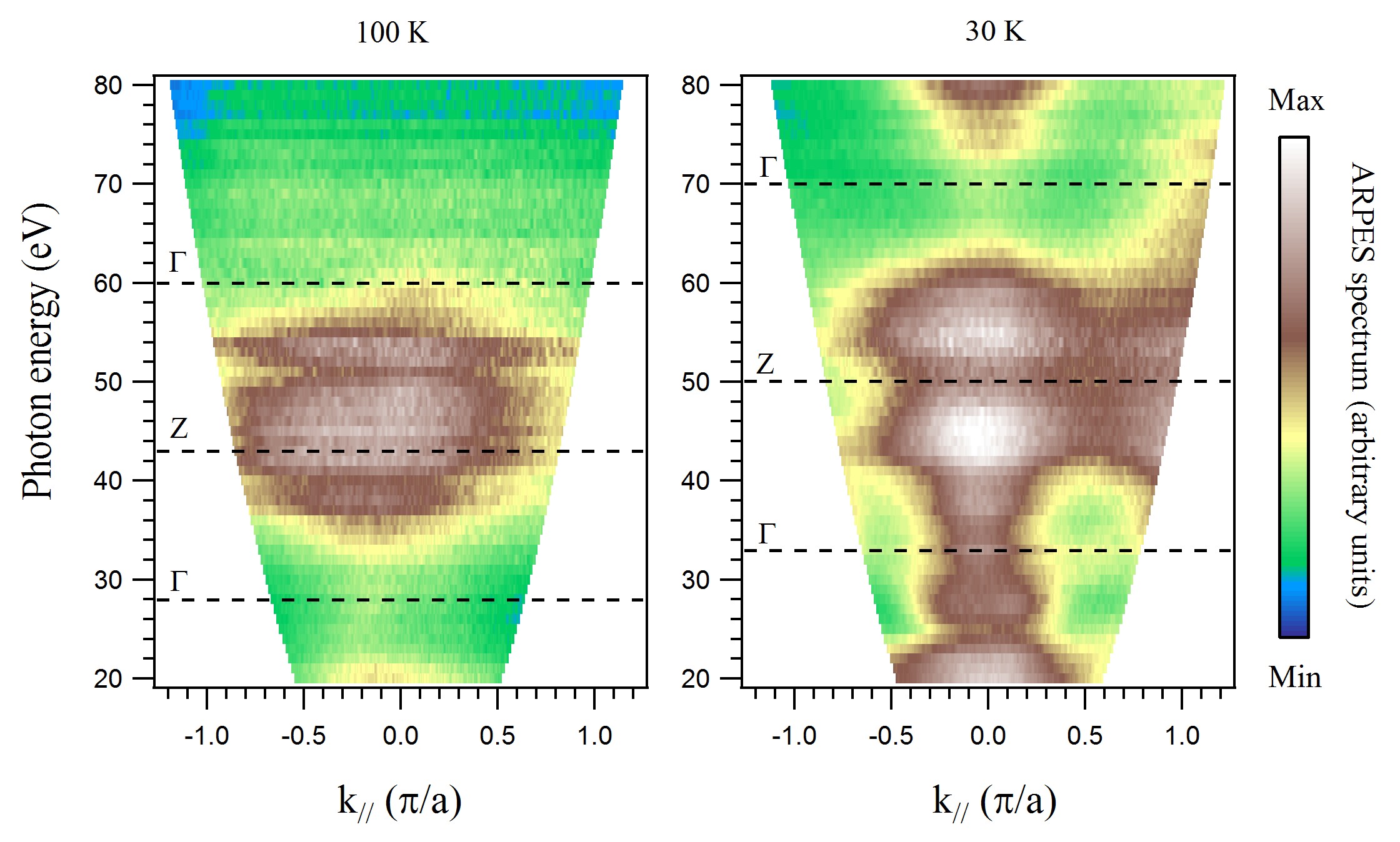} 
\par\end{centering}

\caption{(Color online). Photon energy dependence of the ARPES spectra of CaFe\textsubscript{2}As\textsubscript{2}
in the tetragonal (T = 100 K) and collapsed-tetragonal (T = 30 K)
phases at the Fermi level around the $\Gamma$ point.\label{fig:Photon-energy-dependence}}
\end{figure}

Fig.\ \ref{fig:Photon-energy-dependence} displays the photoemission
spectra centered on the $\Gamma$ point at the Fermi level for different
photon energies, in the collapsed-tetragonal and tetragonal phases.
From the observed periodicity of the spectrum in the collapsed-tetragonal
phase we find that the $\Gamma$ point is located around 33 eV and
70 eV whereas a Z point is found around 50 eV. Using the sudden approximation
and nearly free-electron model for the final state: $k_{\perp}=\sqrt{2m\left(E_{kin}\cos^{2}\theta+V_{0}\right)}/\hbar$
and the lattice parameters of Saparov \emph{et al.} \cite{Saparov-CaFe2As2-collapsed},
we deduce an inner potential $V_{0}$ of about 15 eV, which is consistent
with other Fe-based superconductors \cite{Pierre-ARPES-review}. This
value of the inner potential also corresponds well to the data observed
by Dhaka \emph{et al.} \cite{Dakha-CaFe2As2}. In the tetragonal phase
the data are less clear, but using the same value for $V_{0}$ we
estimate the $\Gamma$ point to be around 28 eV and 60 eV and the
Z point around 43 eV. This assumption is plausible at 28 eV, even
though for higher photon energy there seems to be a slight discrepancy
with the observed spectrum. This might be due to a modification of
the inner potential since the surface will probably be different in
the tetragonal phase.

\begin{figure}
\begin{centering}
\includegraphics[width=8.5cm]{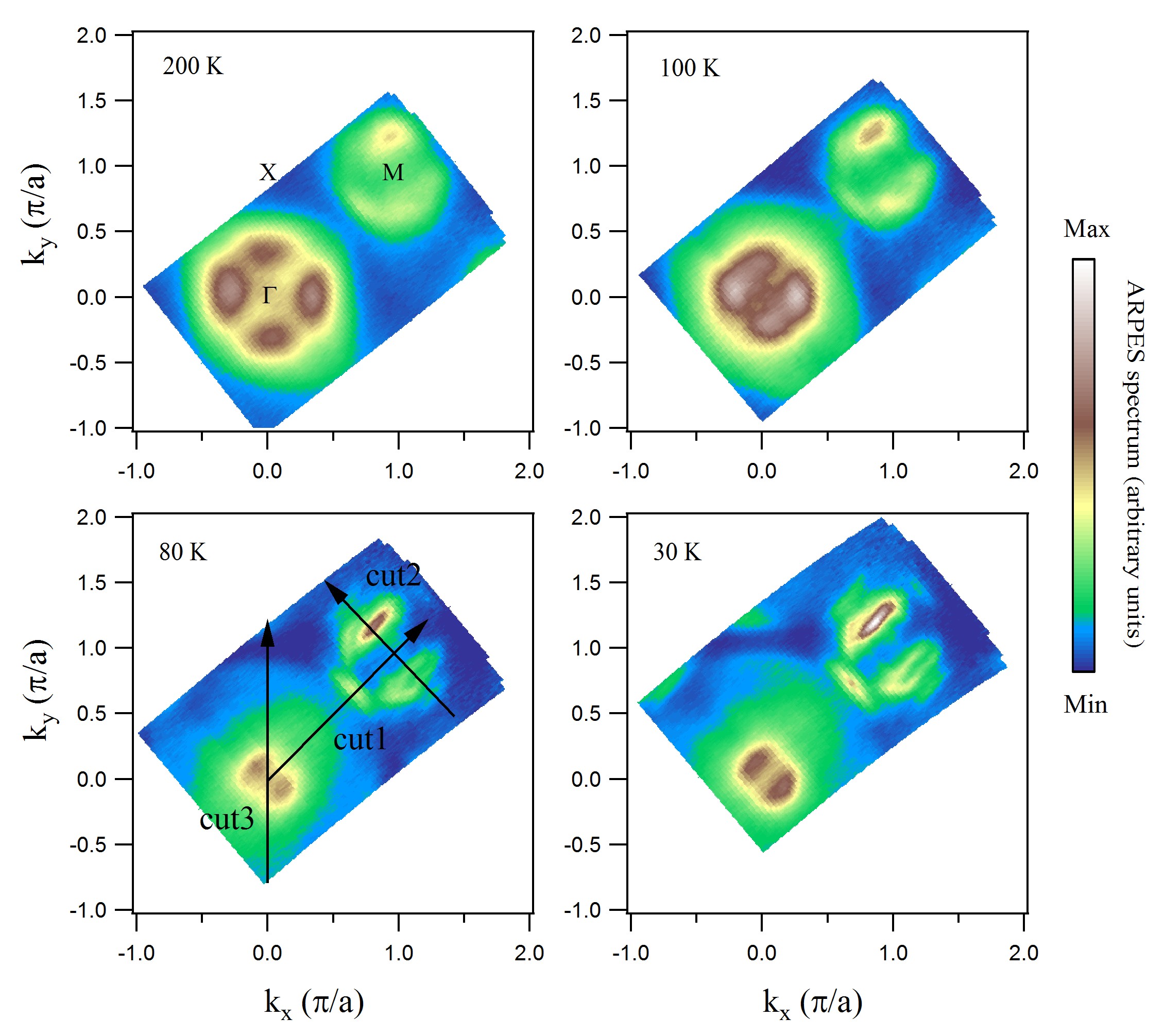} 
\par\end{centering}

\caption{(Color online). Fermi surface mapping of CaFe\textsubscript{2}As\textsubscript{2}
in the tetragonal (T > 90 K) and collapsed-tetragonal (T < 90 K) phases.\label{fig:CaFe2As2-Fermi-surface}}
\end{figure}

Fig.\ \ref{fig:CaFe2As2-Fermi-surface} shows the Fermi surface of
our CaFe\textsubscript{2}As\textsubscript{2} sample in the tetragonal
and collapsed-tetragonal phases recorded with a photon energy of 21.218
eV. We have lowered the temperature from 200 K to 30 K and finished
the measurements less than 30 hours after the cleave, such that the
aging of the sample was not important. Using the previously deduced
inner potential, we find that for the collapsed-tetragonal phase the
$\Gamma$ point %
\footnote{For simplicity, we name the points measured in Fig. \ref{fig:CaFe2As2-Fermi-surface}
as their projection on the $k_{z}=0$ plane.%
} has a $k_{z}$ close to 1.25 $\pi/c'$ -- with $c'=c/2$ the distance
between two FeAs layers, close to the Z point of coordinates $(0,0,\pi/c')$.
$k_{z}$ then decreases when $k_{\parallel}$ is increased, with a
value of 0.89 $\pi/c'$ at the M point and 1.01 $\pi/c'$ at the X
point. It is interesting to note that the point symmetric to $\Gamma$
with respect to the X point would have for coordinates $(2\pi/a,0,0.5\pi/c')$,
such that it would nearly correspond to the same high-symmetry point
\footnote{Indeed, the point Z with coordinates $(0,0,\pi/c')$ is equivalent
to the point with coordinates $(2\pi/a,0,0)$.%
}. For the tetragonal phase, we find $k_{z}=1.6\pi/c'$ at the $\Gamma$
point and $k_{z}=1.19\pi/c'$ at the M point.

\begin{figure}
\begin{centering}
\includegraphics[width=8.5cm]{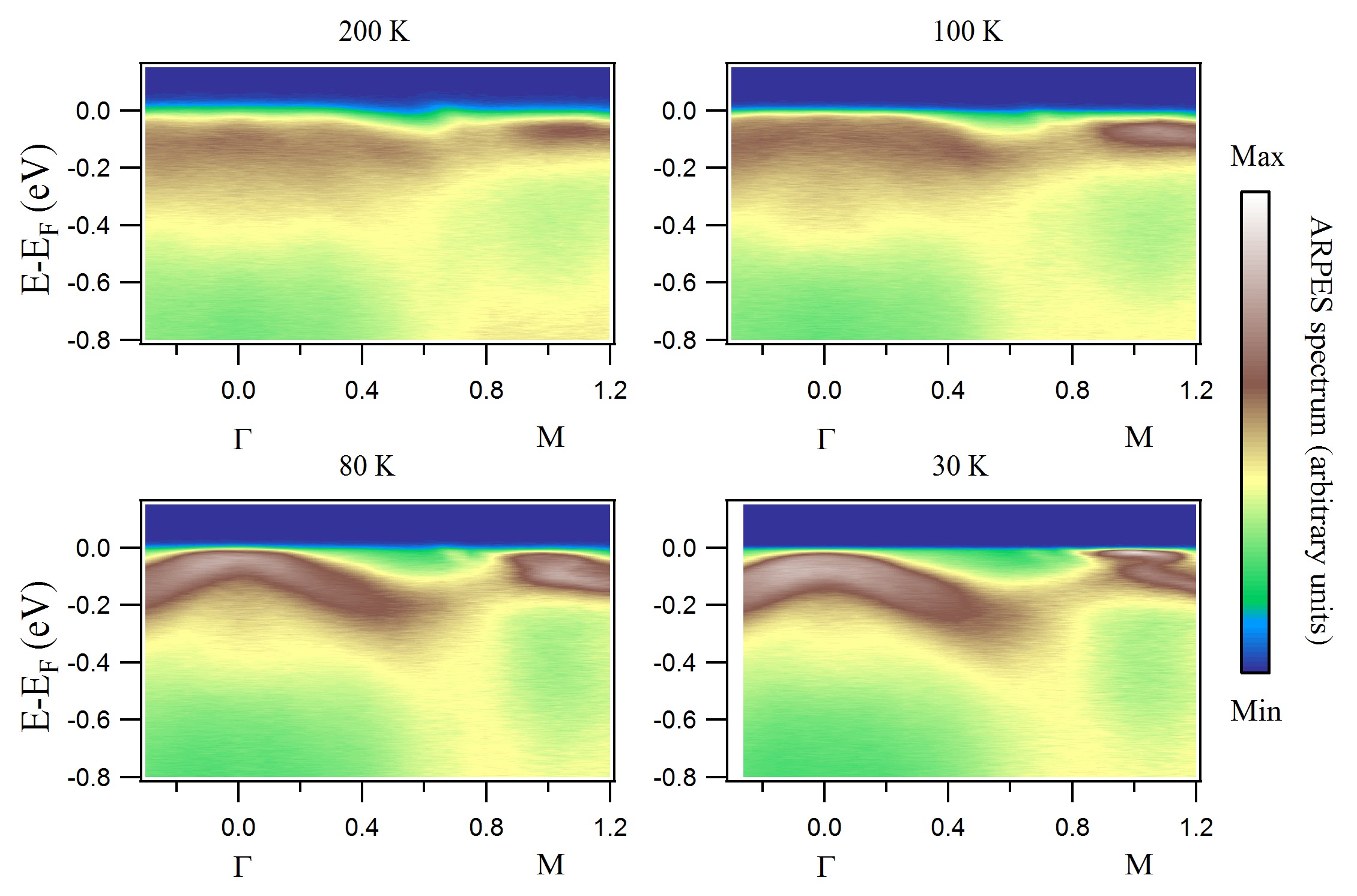} 
\par\end{centering}

\begin{centering}
\includegraphics[width=8.5cm]{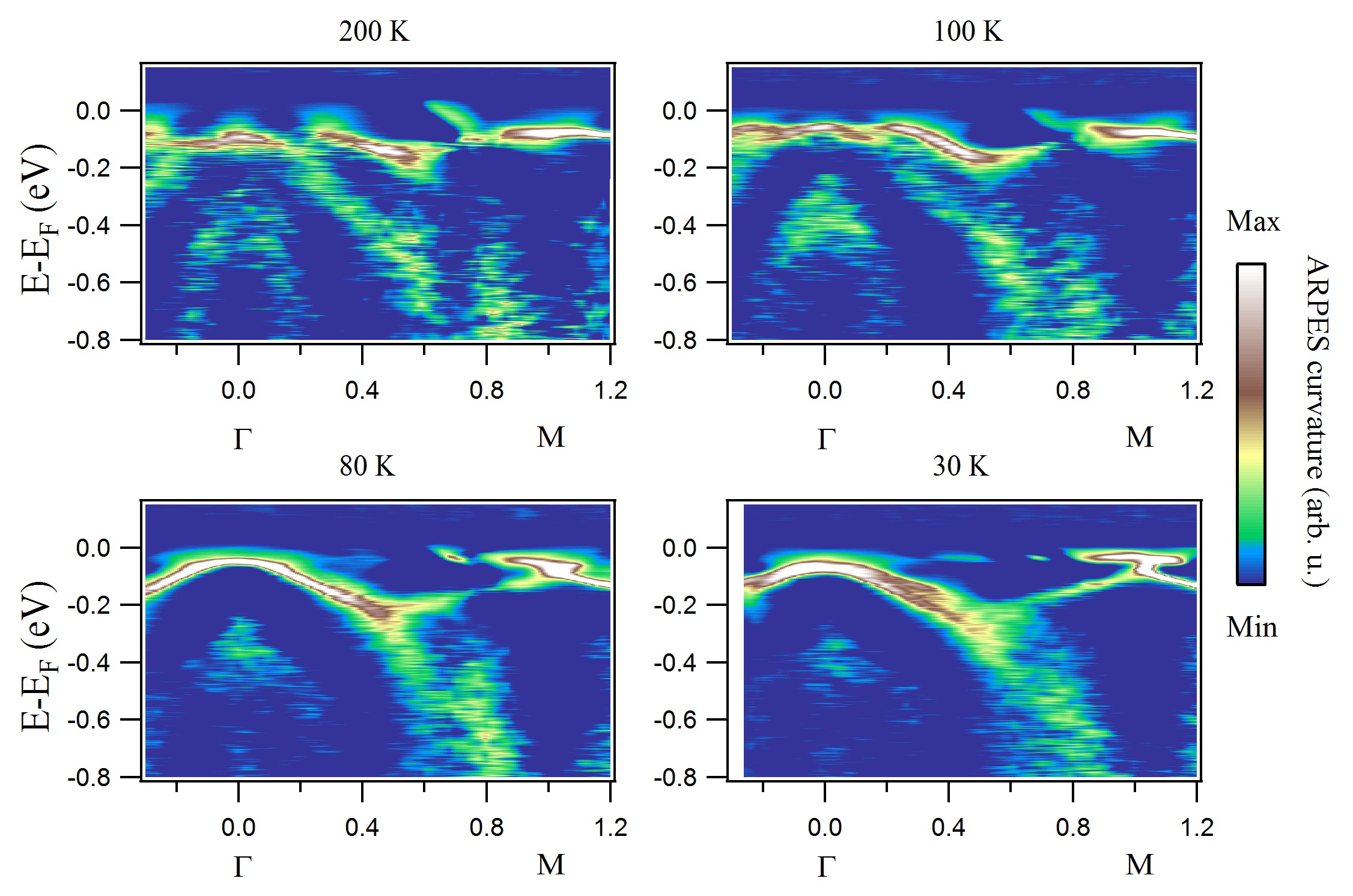} 
\par\end{centering}

\caption{(Color online). ARPES spectra (top) and curvature (bottom) of CaFe\textsubscript{2}As\textsubscript{2}
in the tetragonal (T > 90 K) and collapsed-tetragonal (T < 90 K) phases
along cut 1 of Fig.\ \ref{fig:CaFe2As2-Fermi-surface}.\label{fig:GM-CaFe2As2}}
\end{figure}

\begin{figure}
\begin{centering}
\includegraphics[width=8.5cm]{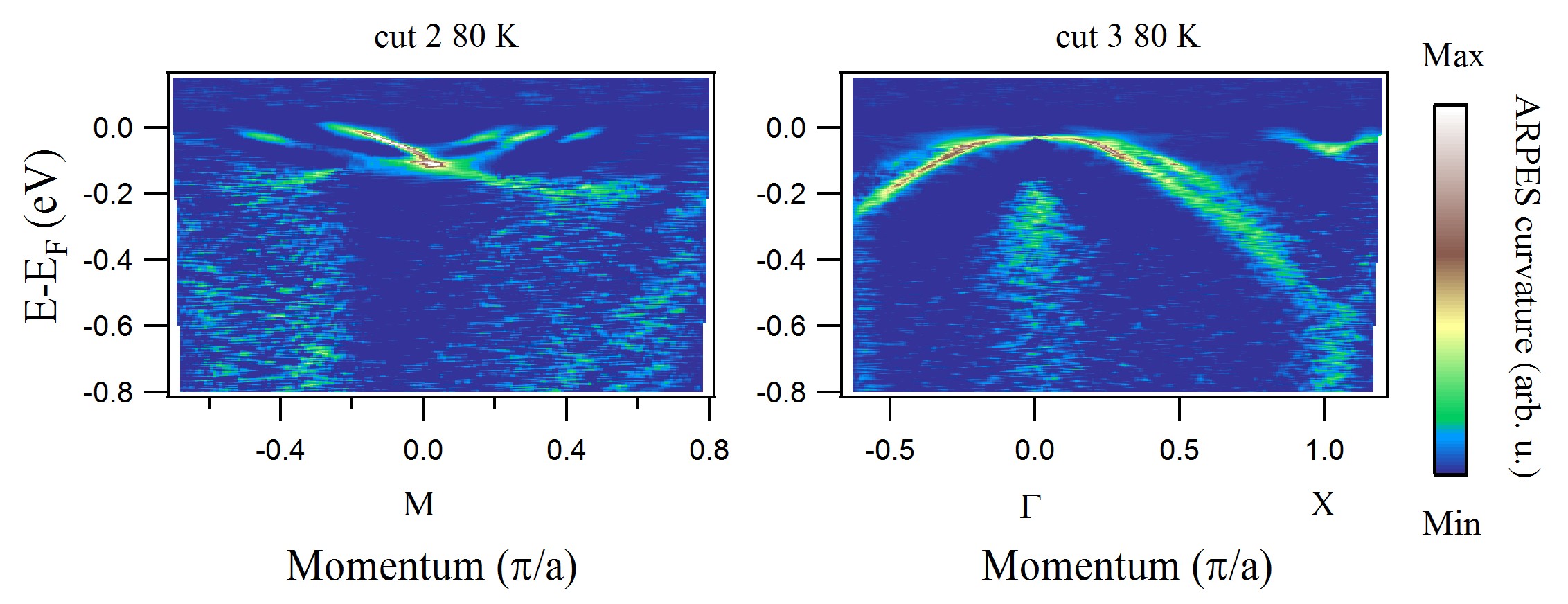} 
\par\end{centering}

\caption{(Color online). Curvature of the ARPES spectra of CaFe\textsubscript{2}As\textsubscript{2}
in the collapsed-tetragonal phase (T < 90 K) along cut 2 (near the
M point, left) and along cut 3 (right) of Fig.\ \ref{fig:CaFe2As2-Fermi-surface}.\label{fig:GX-kperp-curvature-CaFe2As2}}
\end{figure}

For a more detailed analysis of the states forming the Fermi surface,
we also present three different cuts. We first show the $\Gamma$-M
direction for all temperatures (see Fig.\ \ref{fig:GM-CaFe2As2}
for the spectra and its curvature \cite{Peng-curvature}). We also
display a cut near the M point on the direction perpendicular to $\Gamma$-M
(see Fig.\ \ref{fig:GX-kperp-curvature-CaFe2As2} left panel) and
another one along the $\Gamma$-X direction (Fig.\ \ref{fig:GX-kperp-curvature-CaFe2As2}
right panel), for the collapsed-tetragonal phase at 80 K only (similar
results are obtained at 30 K). In the tetragonal phase, we can distinguish
two hole-like bands forming circular hole pockets near the $\Gamma$
point, although one may not cross the Fermi level. We also find two
electron pockets around the M point. This is similar to what is found
in many iron pnictides, and in particular in BaFe\textsubscript{2}As\textsubscript{2}.
Below the transition temperature, the Fermi surface is reorganized.
The circular hole pocket around the $\Gamma$ point shrinks drastically
-- or even disappears -- while a large square hole pocket develops.
This shape of the hole-like bands is very characteristic of the collapsed
structure and qualitatively different from what is seen in BaFe\textsubscript{2}As\textsubscript{2}
\cite{Kaminski-BaFe2As2-CaFe2As2}. This effect is due to a stronger
three-dimensional character, as can be observed from the photon-energy
dependent data of Fig.\ \ref{fig:Photon-energy-dependence}. Indeed,
the $k_{z}$ dispersion is enhanced by the strong As-As $p_{z}$ interlayer
hybridization in the collapsed phase. On the other hand, the electron
pockets near M keep a similar size.

From the temperature-dependent photoemission spectra of Fig.\ \ref{fig:CaFe2As2-Fermi-surface},
it is interesting to see how the features become better defined as
temperature is lowered. Notably, there is a clear difference between
spectra above (at 100 K) and below (at 80 K) the collapse transition.
However, because the quasiparticle dispersions are also changed through
this transition, and because overall the spectrum appears to be very
sensitive to temperature, it is difficult to attribute this improvement
to the transition only.

\section{DFT+DMFT calculations}

We now turn to a theoretical description of the spectral properties
of CaFe\textsubscript{2}As\textsubscript{2}, using \textit{first
principles} dynamical mean field theory (DMFT) techniques. The first
step are calculations based on the by now well-established DFT+DMFT
method \cite{LDA+DMFT-licht,LDA+DMFT-anisimov-1997}. We use the DFT+DMFT
implementation of \cite{cRPA-DMFT-LaOFeAs-markus} within the Local
Density Approximation (LDA) to the exchange-correlation functional,
and Hubbard and Hund's interactions obtained from the constrained
random phase approximation (cRPA) \cite{cRPA-ferdi-2004} in the implementation
of Ref.\ \cite{TMO-vaugier}. The cRPA calculations yield $F^{0}=2.5$
eV, $F^{2}=6.0$ eV and $F^{4}=4.5$ eV, corresponding to a Hund's
rule coupling of $J=0.75$ eV.

\begin{table}
\begin{centering}
\begin{tabular}{ccccc}
\hline 
 & $d_{z^{2}}$ & $d_{x^{2}-y^{2}}$ & $d_{xy}$ & $d_{xz+yz}$\tabularnewline
\hline 
Tetragonal & 1.43 & 1.37 & 1.64 & 1.57\tabularnewline
collapsed-tetragonal & 1.35 & 1.36 & 1.45 & 1.46\tabularnewline
\hline 
\end{tabular}
\par\end{centering}

\caption{Mass renormalizations calculated from DFT+DMFT for the Fe-3\textit{d}
orbitals.\label{tab:Renormalizations}}

\end{table}

\begin{figure}
\begin{centering}
\includegraphics[width=8.5cm]{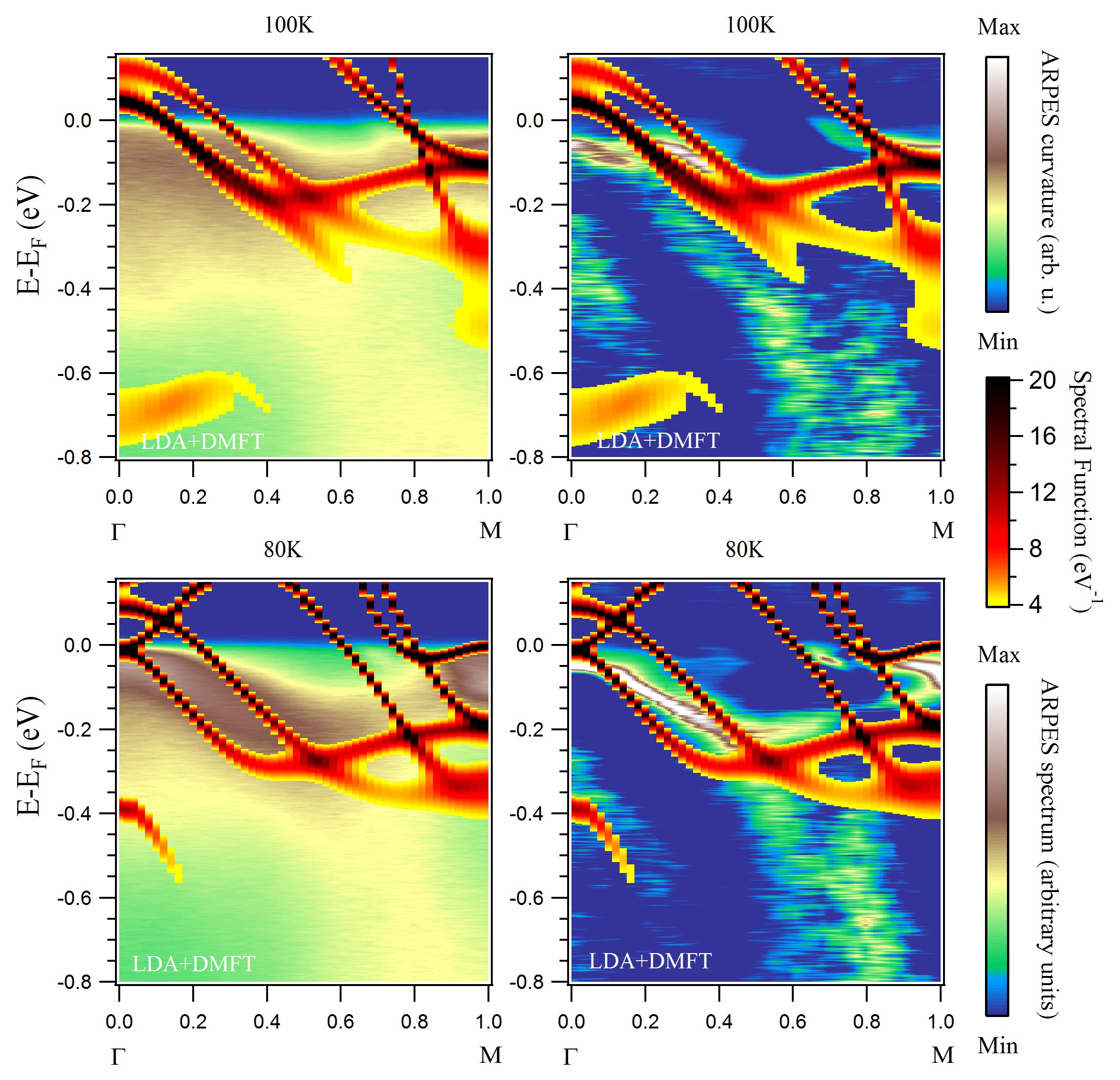} 
\par\end{centering}

\caption{(Color online). Comparison of DFT+DMFT spectral functions with ARPES
spectra of CaFe\textsubscript{2}As\textsubscript{2} in the tetragonal
and collapsed-tetragonal phases. The parts of the spectral functions
with value higher than 4 eV\textsuperscript{-1} are superimposed
on the ARPES data of CaFe\textsubscript{2}As\textsubscript{2} in
the tetragonal (100 K) and collapsed-tetragonal (80 K) phases along
the $\Gamma$-M direction, represented using the (left) spectra or
(right) curvature.\label{fig:LDA+DMFT-CaFe2As2}}
\end{figure}

Fig.\ \ref{fig:LDA+DMFT-CaFe2As2} presents the superposition of
bands extracted from DFT+DMFT calculations performed at 120 K with
the ARPES spectrum along the $\Gamma$-M direction. We have taken
into account the variation of $k_{z}$ as indicated previously. Overall,
the band renormalization is correctly described by the DFT+DMFT calculations.
The theoretical quasi-particle renormalizations as extracted from
a linearization around the Fermi energy of the imaginary part of the
self-energy on the Matsubara axis are displayed in Table \ref{tab:Renormalizations}.
A caveat is however in order since the linear regime is really restricted
to the first few Matsubara frequencies only, indicating that at the
temperature of the calculation the system is at the border to an incoherent
regime. Interestingly, on larger energy scales, at least in the tetragonal
phase the self-energy can quite well be fit as a power law behavior
$\omega^{\alpha}$ with $\alpha$ around 0.75. This is reminiscent
to what was found in BaFe\textsubscript{2}As\textsubscript{2} in
\cite{udyn-werner}.

In agreement with Refs. \cite{Mandal-CaFe2As2,Diehl-CaFe2As2}, we
find the tetragonal phase to exhibit stronger electronic correlations
than the collapsed phase. Within a given phase, we observe stronger
effects on the $d_{xz+yz}$ and $d_{xy}$ orbitals than on the $d_{z^{2}}$
and $d_{x^{2}-y^{2}}$ ones.

We can also see from the calculations that there may be three hole-like
bands in total in the tetragonal phase but two are nearly degenerate
near the $\Gamma$ point. However, if we look at the precise details
of the low-energy states, we can find several discrepancies. In the
tetragonal phase at the Fermi level, one of the bands near the $\Gamma$
point is not well described. It is not clear if this is due to possible
surface effects, limitations of the calculations or other issues.
On the other hand the electron pockets are well described. In the
collapsed-tetragonal phase, the two hole-like bands near the $\Gamma$
point appear to be very close to each other from photoemission measurements,
as can be seen even more clearly on the 30 K data of Fig.\ \ref{fig:GM-CaFe2As2}
and along the $\Gamma$-X direction of Fig.\ \ref{fig:GX-kperp-curvature-CaFe2As2}.
At the M point, two bands are responsible for the electron pockets,
however the shape deviates from the experimental data due to upbending
of one of the bands. On the other hand, we consider the agreement
for the Fermi vector of the large hole pocket relatively satisfying
since this band is very sensitive to the precise value of $k_{z}$.
If we suppose that the ARPES spectrum reflects the bulk features of
the collapsed-tetragonal phase, an important test for improved calculational
schemes will be the correct prediction of the dispersion of the two
hole-like bands near $\Gamma$, and of the interesting topology found
near the M point in Fig.\ \ref{fig:GX-kperp-curvature-CaFe2As2},
which shows three bands crossing the Fermi level very close to each
other -- one of them being the large hole pocket. This last point
is very specific to this compound in the iron pnictides family and
due to the large $k_{z}$ dispersion of the collapsed phase. We will
present results beyond current DFT+DMFT techniques for CaFe$_{2}$As$_{2}$
in section 5 below.

\section{Interplay of structural and electronic properties within DFT+DMFT:
interlayer versus intralayer geometries}

The origin of the reduction of correlations in the collapsed-tetragonal
phase compared to the tetragonal phase is challenging to understand
since both Fe-As and As-As bindings are modified. Indeed, in the collapsed
structure the As-As interlayer binding is much stronger, which should
increase the three-dimensional character of the band structure dispersion.
However, the transition has also another effect on the Fe-As binding
since the $c$ axis collapses so much that the As height to the Fe
plane is reduced. The result is that the Fe-As distance is shortened,
suggesting an enhancement of the hybridization between the As-4\textit{p}
and the Fe-3\textit{d} orbitals -- though the expansion of the $a$
axis limits this enhancement.

To decouple these two effects we have performed DFT+DMFT calculations
on two hypothetical ``hybrid'' compounds. In the first one, we keep
the same angle and distances between atoms within the FeAs layers
as in the tetragonal phase, while the interlayer As-As distance is
that of the collapsed-tetragonal phase. In the other one, we do the
opposite: the layer is that of the collapsed-tetragonal phase and
the interlayer distance is that of the tetragonal phase.

\begin{figure}
\begin{centering}
\includegraphics[width=8.5cm]{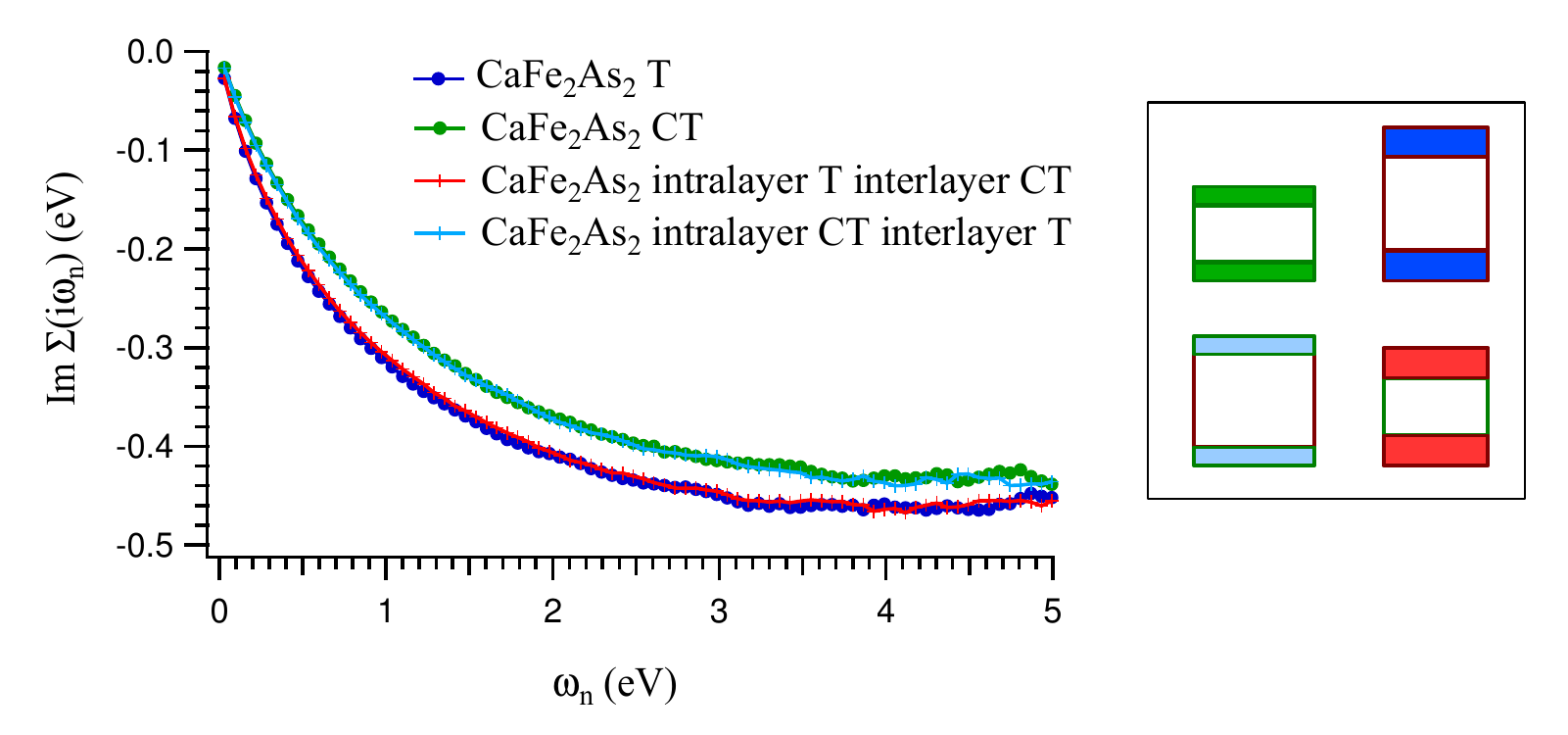} 
\par\end{centering}

\caption{(Color online). Imaginary part of the self-energy in Matsubara frequencies
of the $d_{xy}$ orbital for CaFe\textsubscript{2}As\textsubscript{2}
in the tetragonal structure (T), collapsed-tetragonal structure (CT)
and in two hypothetical structures mixing the interlayer (i.e. As-As
interlayer distance) and intralayer (i.e. intralayer Fe and As angles
and distances) of the tetragonal and collapsed-tetragonal structures.
The inset shows a schematic view of the different structures (colours
correspond to the legend).\label{fig:CaFe2As2-hybrid}}
\end{figure}

The imaginary part of the self-energy of the $d_{xy}$ orbital in
Matsubara frequencies is displayed in Fig.\ \ref{fig:CaFe2As2-hybrid}.
The effect on the $d_{xz}+d_{yz}$ orbital is similar, and since those
orbitals have the highest density-of-states at the Fermi level we
expect that they control the coherence properties of the compound
\footnote{In contrast, there is no difference between the four structures for
the $d_{x^{2}-y^{2}}$ orbital and a much smaller one for the $d_{z^{2}}$
orbital.%
}. We can first see that in the collapsed phase the imaginary part
of the self-energy displays a more coherent behavior, which corresponds
to the longer lifetime of quasiparticles displayed in Fig.\ \ref{fig:LDA+DMFT-CaFe2As2}.
Furthermore, the shape of the self-energy of the hybrid compounds
depends on the structure of the FeAs layer, while it is nearly insensitive
to the interlayer As-As distance. Naturally, in reality those two
effects are linked with each other, since the deformation of the FeAs
layer is caused by the formation of As-As bonds that make the $c$
axis collapse. Still, this numerical experiment indicates that within
DFT+DMFT the improvement of coherence properties is not due to the
interlayer As-As bonding but to the increase of the Fe-As hybridization
within one single layer.

\section{Beyond DFT+DMFT: results from screened-exchange dynamical mean field
theory}

\begin{figure}
\begin{centering}
\includegraphics[width=8.5cm]{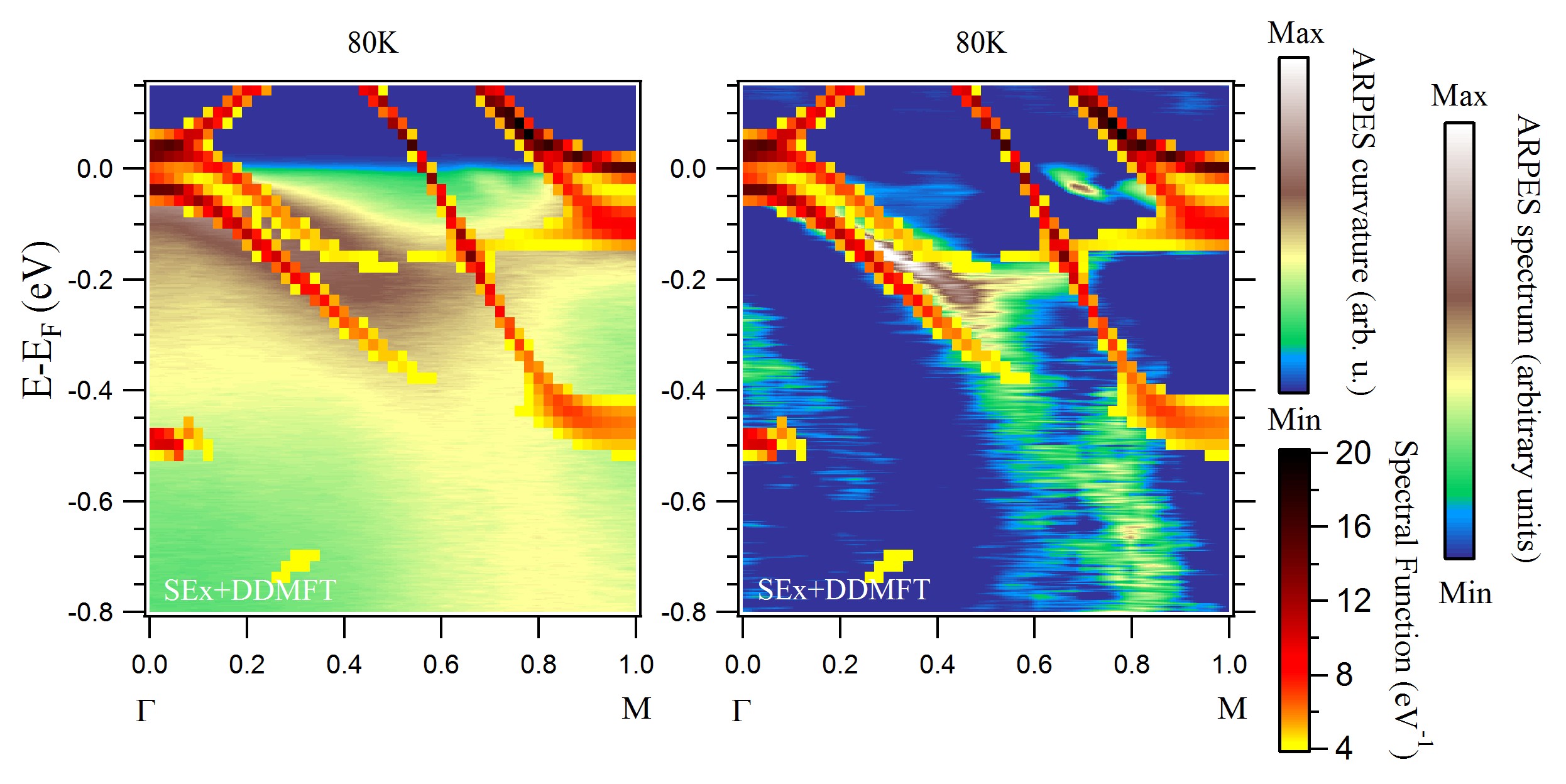} 
\par\end{centering}

\caption{(Color online). Comparison of SEx+DDMFT spectral function with ARPES
spectra of CaFe\textsubscript{2}As\textsubscript{2} in the collapsed-tetragonal
phase. The parts of the spectral functions with value higher than
4 eV\textsuperscript{-1} are superimposed on the ARPES data of CaFe\textsubscript{2}As\textsubscript{2}
along the $\Gamma$-M direction, represented using the (left) spectra
or (right) curvature.\label{fig:SEX+DDMFT}}
\end{figure}

Recently, some of us have proposed a new calculational scheme that
goes beyond current DFT+DMFT techniques \cite{Ambroise-BaCo2As2,Ambroise-SrVO3}:
the combination of a screened exchange Hamiltonian with ``dynamical
DMFT'', that is DMFT extended to dynamical Hubbard interactions \cite{udyn-michele,udyn-werner,SrVO3-dynU-Wang}
was shown to drastically improve upon the low-energy description of
BaCo$_{2}$As$_{2}$. In this context, the apparent success of the
standard DFT+DMFT scheme was shown to result from an error cancelation
effect: a one-body Hamiltonian where the local exchange-correlation
potential of DFT has been replaced by a non-local screened Fock term
has in fact a wider band structure than the DFT one, but including
dynamical screening effects at the level of the Hubbard interactions
leads to additional renormalizations of the electronic states, as
compared to usual DMFT. For this reason, the overall bandwidth of
LDA+DMFT calculations and ``Screened Exchange+Dynamical DMFT'' (``SEx+DDMFT'')
calculations are similar. The low-energy dispersions are however quite
strongly improved by the introduction of the non-local screened exchange
contribution.

In Fig.\ \ref{fig:SEX+DDMFT}, we present the results of this SEx+DDMFT
scheme, in the implementation of Ref.\ \cite{Ambroise-BaCo2As2},
applied to CaFe$_{2}$As$_{2}$. We use the frequency-dependent interactions
as calculated in Ref.\ \cite{Shell-folding} and a value of $\lambda=1.7$
$a_{0}^{-1}$ for the screening wavelength which corresponds to the
density-of-states at the Fermi level of the calculated result. As
anticipated, due to the antagonistic effects of the non-local screened-exchange
and of the high-frequency tail of the interactions, the overall renormalization
of SEx+DDMFT is similar to the DFT+DMFT one. Still, the details of
the quasiparticles dispersions at low energy are importantly modified.
In the collapsed-tetragonal phase of CaFe\textsubscript{2}As\textsubscript{2},
the two bands observed near the $\Gamma$ point are correctly described,
in contrast to DFT+DMFT. The third large hole pocket is at variance
with the calculations, but we stress that its precise Fermi vector
is very sensitive on the value of $k_{z}$.

\section{Conclusion}

We have performed a study of the tetragonal and collapsed-tetragonal
phases of CaFe\textsubscript{2}As\textsubscript{2} using ARPES and
electronic structure calculations. Our results support the picture
that within DMFT, the collapsed-tetragonal phase exhibits reduced
correlations and higher coherence temperature due to the higher Fe-As
hybridization, in agreement with other studies \cite{Mandal-CaFe2As2,Diehl-CaFe2As2}.
However, we note that the reduction of correlations that we observe
does not result in a dramatic change in the electronic self-energy
that could explain by itself the behavior seen by resistivity measurements.
This is confirmed by the ARPES results in the sense that the quasiparticle
lifetimes away from the $\Gamma$ point do not seem to be strongly
impacted by the transition. Around the $\Gamma$ point itself the
electronic states found by photoemission appear more coherent but
their dispersion has been largely reshaped by the collapse of the
crystal. Furthermore, the unconventional transport behavior observed
in experiments might not be the result of a change in electronic coherence
alone. The reconstruction of the Fermi surface and of the low-energy
electronic dispersion might induce geometric effects \emph{via}, e.g.
the Fermi surface nesting or dimensionality. Finally, at the temperature
where the $T$ -- or $T^{1.5}$ -- behavior of the resistivity is
observed, phonons likely also come into play.

\section*{ACKNOWLEDGMENTS}

We acknowledge useful discussions with V\'{e}ronique Brouet and the
coauthors of Ref.\ \cite{Ambroise-BaCo2As2}, in particular Thomas
Ayral, Michel Ferrero and Olivier Parcollet for support on the TRIQS
toolkit \cite{TRIQS-website}. We acknowledge SOLEIL for provision
of synchrotron radiation facilities and we would like to thank Fran\c{c}ois
Bertran, Patrick Le F\`{e}vre and Amina Taleb for assistance in using
the beamline CASSIOPEE. This work was supported by the Cai Yuanpei
program, IDRIS/GENCI Orsay under project 091393 and the European Research
Council under project 617196. We also acknowledge grants from MOST
(2010CB923000 and 2011CBA001000, 2011CBA00102, 2012CB821403) and NSFC
(10974175, 11004232, 11034011/A0402, 11234014 and 11274362) from China.
The work at Oak Ridge National Laboratory was primarily supported
by the U. S. Department of Energy, Office of Science, Basic Energy
Sciences, Materials Science and Engineering Division.

\bibliographystyle{plain}\input{CaFe2As2.bbl}

\begin{thebibliography}{10}%
\makeatletter
\providecommand \@ifxundefined [1]{%
 \ifx #1\undefined \expandafter \@firstoftwo
 \else \expandafter \@secondoftwo
\fi
}%
\providecommand \@ifnum [1]{%
 \ifnum #1\expandafter \@firstoftwo
 \else \expandafter \@secondoftwo
\fi
}%
\providecommand \enquote [1]{``#1''}%
\providecommand \bibnamefont  [1]{#1}%
\providecommand \bibfnamefont [1]{#1}%
\providecommand \citenamefont [1]{#1}%
\providecommand\href[0]{\@sanitize\@href}%
\providecommand\@href[1]{\endgroup\@@startlink{#1}\endgroup\@@href}%
\providecommand\@@href[1]{#1\@@endlink}%
\providecommand \@sanitize [0]{\begingroup\catcode`\&12\catcode`\#12\relax}%
\@ifxundefined \pdfoutput {\@firstoftwo}{%
 \@ifnum{\z@=\pdfoutput}{\@firstoftwo}{\@secondoftwo}%
}{%
 \providecommand\@@startlink[1]{\leavevmode\special{html:<a href="#1">}}%
 \providecommand\@@endlink[0]{\special{html:</a>}}%
}{%
 \providecommand\@@startlink[1]{%
  \leavevmode
  \pdfstartlink
   attr{/Border[0 0 1 ]/H/I/C[0 1 1]}%
   user{/Subtype/Link/A<</Type/Action/S/URI/URI(#1)>>}%
  \relax
 }%
 \providecommand\@@endlink[0]{\pdfendlink}%
}%
\providecommand \url  [0]{\begingroup\@sanitize \@url }%
\providecommand \@url [1]{\endgroup\@href {#1}{\urlprefix}}%
\providecommand \urlprefix [0]{URL }%
\providecommand \Eprint[0]{\href }%
\@ifxundefined \urlstyle {%
  \providecommand \doi [1]{doi:\discretionary{}{}{}#1}%
}{%
  \providecommand \doi [0]{doi:\discretionary{}{}{}\begingroup
  \urlstyle{rm}\Url }%
}%
\providecommand \doibase [0]{http://dx.doi.org/}%
\providecommand \Doi[1]{\href{\doibase#1}}%
\providecommand \bibAnnote [3]{%
  \BibitemShut{#1}%
  \begin{quotation}\noindent
    \textsc{Key:}\ #2\\\textsc{Annotation:}\ #3%
  \end{quotation}%
}%
\providecommand \bibAnnoteFile [2]{%
  \IfFileExists{#2}{\bibAnnote {#1} {#2} {\input{#2}}}{}%
}%
\providecommand \typeout [0]{\immediate \write \m@ne }%
\providecommand \selectlanguage [0]{\@gobble}%
\providecommand \bibinfo [0]{\@secondoftwo}%
\providecommand \bibfield [0]{\@secondoftwo}%
\providecommand \translation [1]{[#1]}%
\providecommand \BibitemOpen[0]{}%
\providecommand \bibitemStop [0]{}%
\providecommand \bibitemNoStop [0]{.\EOS\space}%
\providecommand \EOS [0]{\spacefactor3000\relax}%
\providecommand \BibitemShut [1]{\csname bibitem#1\endcsname}%
\bibitem{Ronning-CaFe2As2}%
  \BibitemOpen
  \bibfield{author}{%
  \bibinfo {author} {\bibfnamefont{F.}~\bibnamefont{Ronning}}, \bibinfo
  {author} {\bibfnamefont{T.}~\bibnamefont{Klimczuk}}, \bibinfo {author}
  {\bibfnamefont{E.~D.}\ \bibnamefont{Bauer}}, \bibinfo {author}
  {\bibfnamefont{H.}~\bibnamefont{Volz}},\ and\ \bibinfo {author}
  {\bibfnamefont{J.~D.}\ \bibnamefont{Thompson}},\ }%
  \bibfield{journal}{%
  \bibinfo {journal} {J. Phys.: Condens. Matter}\ }%
  \textbf{\bibinfo {volume} {20}},\ \bibinfo {pages} {322201} (\bibinfo {year}
  {2008})%
  \bibAnnoteFile{NoStop}{Ronning-CaFe2As2}%
\bibitem{Ni-CaFe2As2}%
  \BibitemOpen
  \bibfield{author}{%
  \bibinfo {author} {\bibfnamefont{N.}~\bibnamefont{Ni}}, \bibinfo {author}
  {\bibfnamefont{S.}~\bibnamefont{Nandi}}, \bibinfo {author}
  {\bibfnamefont{A.}~\bibnamefont{Kreyssig}}, \bibinfo {author}
  {\bibfnamefont{A.~I.}\ \bibnamefont{Goldman}}, \bibinfo {author}
  {\bibfnamefont{E.~D.}\ \bibnamefont{Mun}}, \bibinfo {author}
  {\bibfnamefont{S.~L.}\ \bibnamefont{Bud'ko}},\ and\ \bibinfo {author}
  {\bibfnamefont{P.~C.}\ \bibnamefont{Canfield}},\ }%
  \bibfield{journal}{%
  \bibinfo {journal} {Phys. Rev. B}\ }%
  \textbf{\bibinfo {volume} {78}},\ \bibinfo {pages} {014523} (\bibinfo {month}
  {Jul}\ \bibinfo {year} {2008})%
  \bibAnnoteFile{NoStop}{Ni-CaFe2As2}%
\bibitem{Diallo-CaFe2As2-spin}%
  \BibitemOpen
  \bibfield{author}{%
  \bibinfo {author} {\bibfnamefont{S.~O.}\ \bibnamefont{Diallo}}, \bibinfo
  {author} {\bibfnamefont{V.~P.}\ \bibnamefont{Antropov}}, \bibinfo {author}
  {\bibfnamefont{T.~G.}\ \bibnamefont{Perring}}, \bibinfo {author}
  {\bibfnamefont{C.}~\bibnamefont{Broholm}}, \bibinfo {author}
  {\bibfnamefont{J.~J.}\ \bibnamefont{Pulikkotil}}, \bibinfo {author}
  {\bibfnamefont{N.}~\bibnamefont{Ni}}, \bibinfo {author}
  {\bibfnamefont{S.~L.}\ \bibnamefont{Bud'ko}}, \bibinfo {author}
  {\bibfnamefont{P.~C.}\ \bibnamefont{Canfield}}, \bibinfo {author}
  {\bibfnamefont{A.}~\bibnamefont{Kreyssig}}, \bibinfo {author}
  {\bibfnamefont{A.~I.}\ \bibnamefont{Goldman}},\ and\ \bibinfo {author}
  {\bibfnamefont{R.~J.}\ \bibnamefont{McQueeney}},\ }%
  \bibfield{journal}{%
  \bibinfo {journal} {Phys. Rev. Lett.}\ }%
  \textbf{\bibinfo {volume} {102}},\ \bibinfo {pages} {187206} (\bibinfo
  {month} {May}\ \bibinfo {year} {2009})%
  \bibAnnoteFile{NoStop}{Diallo-CaFe2As2-spin}%
\bibitem{Torikachvili-CaFe2As2-SC}%
  \BibitemOpen
  \bibfield{author}{%
  \bibinfo {author} {\bibfnamefont{M.~S.}\ \bibnamefont{Torikachvili}},
  \bibinfo {author} {\bibfnamefont{S.~L.}\ \bibnamefont{Bud'ko}}, \bibinfo
  {author} {\bibfnamefont{N.}~\bibnamefont{Ni}},\ and\ \bibinfo {author}
  {\bibfnamefont{P.~C.}\ \bibnamefont{Canfield}},\ }%
  \bibfield{journal}{%
  \bibinfo {journal} {Phys. Rev. Lett.}\ }%
  \textbf{\bibinfo {volume} {101}},\ \bibinfo {pages} {057006} (\bibinfo
  {month} {Jul}\ \bibinfo {year} {2008})%
  \bibAnnoteFile{NoStop}{Torikachvili-CaFe2As2-SC}%
\bibitem{Kreyssig-CaFe2As2-collapsed}%
  \BibitemOpen
  \bibfield{author}{%
  \bibinfo {author} {\bibfnamefont{A.}~\bibnamefont{Kreyssig}}, \bibinfo
  {author} {\bibfnamefont{M.~A.}\ \bibnamefont{Green}}, \bibinfo {author}
  {\bibfnamefont{Y.}~\bibnamefont{Lee}}, \bibinfo {author}
  {\bibfnamefont{G.~D.}\ \bibnamefont{Samolyuk}}, \bibinfo {author}
  {\bibfnamefont{P.}~\bibnamefont{Zajdel}}, \bibinfo {author}
  {\bibfnamefont{J.~W.}\ \bibnamefont{Lynn}}, \bibinfo {author}
  {\bibfnamefont{S.~L.}\ \bibnamefont{Bud'ko}}, \bibinfo {author}
  {\bibfnamefont{M.~S.}\ \bibnamefont{Torikachvili}}, \bibinfo {author}
  {\bibfnamefont{N.}~\bibnamefont{Ni}}, \bibinfo {author}
  {\bibfnamefont{S.}~\bibnamefont{Nandi}}, \bibinfo {author}
  {\bibfnamefont{J.~B.}\ \bibnamefont{Le\~ao}}, \bibinfo {author}
  {\bibfnamefont{S.~J.}\ \bibnamefont{Poulton}}, \bibinfo {author}
  {\bibfnamefont{D.~N.}\ \bibnamefont{Argyriou}}, \bibinfo {author}
  {\bibfnamefont{B.~N.}\ \bibnamefont{Harmon}}, \bibinfo {author}
  {\bibfnamefont{R.~J.}\ \bibnamefont{McQueeney}}, \bibinfo {author}
  {\bibfnamefont{P.~C.}\ \bibnamefont{Canfield}},\ and\ \bibinfo {author}
  {\bibfnamefont{A.~I.}\ \bibnamefont{Goldman}},\ }%
  \bibfield{journal}{%
  \bibinfo {journal} {Phys. Rev. B}\ }%
  \textbf{\bibinfo {volume} {78}},\ \bibinfo {pages} {184517} (\bibinfo {month}
  {Nov}\ \bibinfo {year} {2008})%
  \bibAnnoteFile{NoStop}{Kreyssig-CaFe2As2-collapsed}%
\bibitem{Ran-CaFe2As2-collapsed}%
  \BibitemOpen
  \bibfield{author}{%
  \bibinfo {author} {\bibfnamefont{S.}~\bibnamefont{Ran}}, \bibinfo {author}
  {\bibfnamefont{S.~L.}\ \bibnamefont{Bud'ko}}, \bibinfo {author}
  {\bibfnamefont{D.~K.}\ \bibnamefont{Pratt}}, \bibinfo {author}
  {\bibfnamefont{A.}~\bibnamefont{Kreyssig}}, \bibinfo {author}
  {\bibfnamefont{M.~G.}\ \bibnamefont{Kim}}, \bibinfo {author}
  {\bibfnamefont{M.~J.}\ \bibnamefont{Kramer}}, \bibinfo {author}
  {\bibfnamefont{D.~H.}\ \bibnamefont{Ryan}}, \bibinfo {author}
  {\bibfnamefont{W.~N.}\ \bibnamefont{Rowan-Weetaluktuk}}, \bibinfo {author}
  {\bibfnamefont{Y.}~\bibnamefont{Furukawa}}, \bibinfo {author}
  {\bibfnamefont{B.}~\bibnamefont{Roy}}, \bibinfo {author}
  {\bibfnamefont{A.~I.}\ \bibnamefont{Goldman}},\ and\ \bibinfo {author}
  {\bibfnamefont{P.~C.}\ \bibnamefont{Canfield}},\ }%
  \bibfield{journal}{%
  \bibinfo {journal} {Phys. Rev. B}\ }%
  \textbf{\bibinfo {volume} {83}},\ \bibinfo {pages} {144517} (\bibinfo {month}
  {Apr}\ \bibinfo {year} {2011})%
  \bibAnnoteFile{NoStop}{Ran-CaFe2As2-collapsed}%
\bibitem{Saparov-CaFe2As2-collapsed}%
  \BibitemOpen
  \bibfield{author}{%
  \bibinfo {author} {\bibfnamefont{B.}~\bibnamefont{Saparov}}, \bibinfo
  {author} {\bibfnamefont{C.}~\bibnamefont{Cantoni}}, \bibinfo {author}
  {\bibfnamefont{M.}~\bibnamefont{Pan}}, \bibinfo {author}
  {\bibfnamefont{T.~C.}\ \bibnamefont{Hogan}}, \bibinfo {author}
  {\bibfnamefont{W.}~\bibnamefont{Ratcliff~II}}, \bibinfo {author}
  {\bibfnamefont{S.~D.}\ \bibnamefont{Wilson}}, \bibinfo {author}
  {\bibfnamefont{K.}~\bibnamefont{Fritsch}}, \bibinfo {author}
  {\bibfnamefont{B.~D.}\ \bibnamefont{Gaulin}},\ and\ \bibinfo {author}
  {\bibfnamefont{A.~S.}\ \bibnamefont{Sefat}},\ }%
  \bibfield{journal}{%
  \bibinfo {journal} {Sci. Rep.}\ }%
  \textbf{\bibinfo {volume} {4}} (\bibinfo {year} {2014})%
  \bibAnnoteFile{NoStop}{Saparov-CaFe2As2-collapsed}%
\bibitem{Coldea-CaFe2P2}%
  \BibitemOpen
  \bibfield{author}{%
  \bibinfo {author} {\bibfnamefont{A.~I.}\ \bibnamefont{Coldea}}, \bibinfo
  {author} {\bibfnamefont{C.~M.~J.}\ \bibnamefont{Andrew}}, \bibinfo {author}
  {\bibfnamefont{J.~G.}\ \bibnamefont{Analytis}}, \bibinfo {author}
  {\bibfnamefont{R.~D.}\ \bibnamefont{McDonald}}, \bibinfo {author}
  {\bibfnamefont{A.~F.}\ \bibnamefont{Bangura}}, \bibinfo {author}
  {\bibfnamefont{J.-H.}\ \bibnamefont{Chu}}, \bibinfo {author}
  {\bibfnamefont{I.~R.}\ \bibnamefont{Fisher}},\ and\ \bibinfo {author}
  {\bibfnamefont{A.}~\bibnamefont{Carrington}},\ }%
  \bibfield{journal}{%
  \bibinfo {journal} {Phys. Rev. Lett.}\ }%
  \textbf{\bibinfo {volume} {103}},\ \bibinfo {pages} {026404} (\bibinfo
  {month} {Jul}\ \bibinfo {year} {2009})%
  \bibAnnoteFile{NoStop}{Coldea-CaFe2P2}%
\bibitem{Danura-Ca(FeRh)2As2}%
  \BibitemOpen
  \bibfield{author}{%
  \bibinfo {author} {\bibfnamefont{M.}~\bibnamefont{Danura}}, \bibinfo {author}
  {\bibfnamefont{K.}~\bibnamefont{Kudo}}, \bibinfo {author}
  {\bibfnamefont{Y.}~\bibnamefont{Oshiro}}, \bibinfo {author}
  {\bibfnamefont{S.}~\bibnamefont{Araki}}, \bibinfo {author}
  {\bibfnamefont{T.~C.}\ \bibnamefont{Kobayashi}},\ and\ \bibinfo {author}
  {\bibfnamefont{M.}~\bibnamefont{Nohara}},\ }%
  \bibfield{journal}{%
  \bibinfo {journal} {J. Phys. Soc. Jpn.}\ }%
  \textbf{\bibinfo {volume} {80}},\ \bibinfo {pages} {103701} (\bibinfo {year}
  {2011})%
  \bibAnnoteFile{NoStop}{Danura-Ca(FeRh)2As2}%
\bibitem{Saha-rare-earth-CaFe2As2}%
  \BibitemOpen
  \bibfield{author}{%
  \bibinfo {author} {\bibfnamefont{S.~R.}\ \bibnamefont{Saha}}, \bibinfo
  {author} {\bibfnamefont{N.~P.}\ \bibnamefont{Butch}}, \bibinfo {author}
  {\bibfnamefont{T.}~\bibnamefont{Drye}}, \bibinfo {author}
  {\bibfnamefont{J.}~\bibnamefont{Magill}}, \bibinfo {author}
  {\bibfnamefont{S.}~\bibnamefont{Ziemak}}, \bibinfo {author}
  {\bibfnamefont{K.}~\bibnamefont{Kirshenbaum}}, \bibinfo {author}
  {\bibfnamefont{P.~Y.}\ \bibnamefont{Zavalij}}, \bibinfo {author}
  {\bibfnamefont{J.~W.}\ \bibnamefont{Lynn}},\ and\ \bibinfo {author}
  {\bibfnamefont{J.}~\bibnamefont{Paglione}},\ }%
  \bibfield{journal}{%
  \bibinfo {journal} {Phys. Rev. B}\ }%
  \textbf{\bibinfo {volume} {85}},\ \bibinfo {pages} {024525} (\bibinfo {month}
  {Jan}\ \bibinfo {year} {2012})%
  \bibAnnoteFile{NoStop}{Saha-rare-earth-CaFe2As2}%
\bibitem{Tsubota-Ca(FeRh)2As2}%
  \BibitemOpen
  \bibfield{author}{%
  \bibinfo {author} {\bibfnamefont{K.}~\bibnamefont{Tsubota}}, \bibinfo
  {author} {\bibfnamefont{T.}~\bibnamefont{Wakita}}, \bibinfo {author}
  {\bibfnamefont{H.}~\bibnamefont{Nagao}}, \bibinfo {author}
  {\bibfnamefont{C.}~\bibnamefont{Hiramatsu}}, \bibinfo {author}
  {\bibfnamefont{T.}~\bibnamefont{Ishiga}}, \bibinfo {author}
  {\bibfnamefont{M.}~\bibnamefont{Sunagawa}}, \bibinfo {author}
  {\bibfnamefont{K.}~\bibnamefont{Ono}}, \bibinfo {author}
  {\bibfnamefont{H.}~\bibnamefont{Kumigashira}}, \bibinfo {author}
  {\bibfnamefont{M.}~\bibnamefont{Danura}}, \bibinfo {author}
  {\bibfnamefont{K.}~\bibnamefont{Kudo}}, \bibinfo {author}
  {\bibfnamefont{M.}~\bibnamefont{Nohara}}, \bibinfo {author}
  {\bibfnamefont{Y.}~\bibnamefont{Muraoka}},\ and\ \bibinfo {author}
  {\bibfnamefont{T.}~\bibnamefont{Yokoya}},\ }%
  \bibfield{journal}{%
  \bibinfo {journal} {J. Phys. Soc. Jpn.}\ }%
  \textbf{\bibinfo {volume} {82}},\ \bibinfo {pages} {073705} (\bibinfo {year}
  {2013})%
  \bibAnnoteFile{NoStop}{Tsubota-Ca(FeRh)2As2}%
\bibitem{Dakha-CaFe2As2}%
  \BibitemOpen
  \bibfield{author}{%
  \bibinfo {author} {\bibfnamefont{R.~S.}\ \bibnamefont{Dhaka}}, \bibinfo
  {author} {\bibfnamefont{R.}~\bibnamefont{Jiang}}, \bibinfo {author}
  {\bibfnamefont{S.}~\bibnamefont{Ran}}, \bibinfo {author}
  {\bibfnamefont{S.~L.}\ \bibnamefont{Bud'ko}}, \bibinfo {author}
  {\bibfnamefont{P.~C.}\ \bibnamefont{Canfield}}, \bibinfo {author}
  {\bibfnamefont{B.~N.}\ \bibnamefont{Harmon}}, \bibinfo {author}
  {\bibfnamefont{A.}~\bibnamefont{Kaminski}}, \bibinfo {author}
  {\bibfnamefont{M.}~\bibnamefont{Tomi\'{c}}}, \bibinfo {author}
  {\bibfnamefont{R.}~\bibnamefont{Valent\'{\i}}},\ and\ \bibinfo {author}
  {\bibfnamefont{Y.}~\bibnamefont{Lee}},\ }%
  \bibfield{journal}{%
  \bibinfo {journal} {Phys. Rev. B}\ }%
  \textbf{\bibinfo {volume} {89}},\ \bibinfo {pages} {020511} (\bibinfo {month}
  {Jan}\ \bibinfo {year} {2014})%
  \bibAnnoteFile{NoStop}{Dakha-CaFe2As2}%
\bibitem{Gofryk-CaFe2As2}%
  \BibitemOpen
  \bibfield{author}{%
  \bibinfo {author} {\bibfnamefont{K.}~\bibnamefont{Gofryk}}, \bibinfo {author}
  {\bibfnamefont{B.}~\bibnamefont{Saparov}}, \bibinfo {author}
  {\bibfnamefont{T.}~\bibnamefont{Durakiewicz}}, \bibinfo {author}
  {\bibfnamefont{A.}~\bibnamefont{Chikina}}, \bibinfo {author}
  {\bibfnamefont{S.}~\bibnamefont{Danzenb\"acher}}, \bibinfo {author}
  {\bibfnamefont{D.~V.}\ \bibnamefont{Vyalikh}}, \bibinfo {author}
  {\bibfnamefont{M.~J.}\ \bibnamefont{Graf}},\ and\ \bibinfo {author}
  {\bibfnamefont{A.~S.}\ \bibnamefont{Sefat}},\ }%
  \bibfield{journal}{%
  \bibinfo {journal} {Phys. Rev. Lett.}\ }%
  \textbf{\bibinfo {volume} {112}},\ \bibinfo {pages} {186401} (\bibinfo
  {month} {May}\ \bibinfo {year} {2014})%
  \bibAnnoteFile{NoStop}{Gofryk-CaFe2As2}%
\bibitem{Yildirim-spin-As}%
  \BibitemOpen
  \bibfield{author}{%
  \bibinfo {author} {\bibfnamefont{T.}~\bibnamefont{Yildirim}},\ }%
  \bibfield{journal}{%
  \bibinfo {journal} {Phys. Rev. Lett.}\ }%
  \textbf{\bibinfo {volume} {102}},\ \bibinfo {pages} {037003} (\bibinfo
  {month} {Jan}\ \bibinfo {year} {2009})%
  \bibAnnoteFile{NoStop}{Yildirim-spin-As}%
\bibitem{Tomic-CaFe2As2}%
  \BibitemOpen
  \bibfield{author}{%
  \bibinfo {author} {\bibfnamefont{M.}~\bibnamefont{Tomi\'{c}}}, \bibinfo
  {author} {\bibfnamefont{R.}~\bibnamefont{Valent\'{\i}}},\ and\ \bibinfo
  {author} {\bibfnamefont{H.~O.}\ \bibnamefont{Jeschke}},\ }%
  \bibfield{journal}{%
  \bibinfo {journal} {Phys. Rev. B}\ }%
  \textbf{\bibinfo {volume} {85}},\ \bibinfo {pages} {094105} (\bibinfo {month}
  {Mar}\ \bibinfo {year} {2012})%
  \bibAnnoteFile{NoStop}{Tomic-CaFe2As2}%
\bibitem{BaCo2As2-Dakha}%
  \BibitemOpen
  \bibfield{author}{%
  \bibinfo {author} {\bibfnamefont{R.~S.}\ \bibnamefont{Dhaka}}, \bibinfo
  {author} {\bibfnamefont{Y.}~\bibnamefont{Lee}}, \bibinfo {author}
  {\bibfnamefont{V.~K.}\ \bibnamefont{Anand}}, \bibinfo {author}
  {\bibfnamefont{D.~C.}\ \bibnamefont{Johnston}}, \bibinfo {author}
  {\bibfnamefont{B.~N.}\ \bibnamefont{Harmon}},\ and\ \bibinfo {author}
  {\bibfnamefont{A.}~\bibnamefont{Kaminski}},\ }%
  \bibfield{journal}{%
  \bibinfo {journal} {Phys. Rev. B}\ }%
  \textbf{\bibinfo {volume} {87}},\ \bibinfo {pages} {214516} (\bibinfo {year}
  {2013})%
  \bibAnnoteFile{NoStop}{BaCo2As2-Dakha}%
\bibitem{Mandal-CaFe2As2}%
  \BibitemOpen
  \bibfield{author}{%
  \bibinfo {author} {\bibfnamefont{S.}~\bibnamefont{Mandal}}, \bibinfo {author}
  {\bibfnamefont{R.~E.}\ \bibnamefont{Cohen}},\ and\ \bibinfo {author}
  {\bibfnamefont{K.}~\bibnamefont{Haule}},\ }%
  \bibfield{journal}{%
  \bibinfo {journal} {Phys. Rev. B}\ }%
  \textbf{\bibinfo {volume} {90}},\ \bibinfo {pages} {060501} (\bibinfo {month}
  {Aug}\ \bibinfo {year} {2014})%
  \bibAnnoteFile{NoStop}{Mandal-CaFe2As2}%
\bibitem{Diehl-CaFe2As2}%
  \BibitemOpen
  \bibfield{author}{%
  \bibinfo {author} {\bibfnamefont{J.}~\bibnamefont{Diehl}}, \bibinfo {author}
  {\bibfnamefont{S.}~\bibnamefont{Backes}}, \bibinfo {author}
  {\bibfnamefont{D.}~\bibnamefont{Guterding}}, \bibinfo {author}
  {\bibfnamefont{H.~O.}\ \bibnamefont{Jeschke}},\ and\ \bibinfo {author}
  {\bibfnamefont{R.}~\bibnamefont{Valent\'{\i}}},\ }%
  \bibfield{journal}{%
  \bibinfo {journal} {Phys. Rev. B}\ }%
  \textbf{\bibinfo {volume} {90}},\ \bibinfo {pages} {085110} (\bibinfo {month}
  {Aug}\ \bibinfo {year} {2014})%
  \bibAnnoteFile{NoStop}{Diehl-CaFe2As2}%
\bibitem{Kasahara-Fermi-liquid-CaFe2As2}%
  \BibitemOpen
  \bibfield{author}{%
  \bibinfo {author} {\bibfnamefont{S.}~\bibnamefont{Kasahara}}, \bibinfo
  {author} {\bibfnamefont{T.}~\bibnamefont{Shibauchi}}, \bibinfo {author}
  {\bibfnamefont{K.}~\bibnamefont{Hashimoto}}, \bibinfo {author}
  {\bibfnamefont{Y.}~\bibnamefont{Nakai}}, \bibinfo {author}
  {\bibfnamefont{H.}~\bibnamefont{Ikeda}}, \bibinfo {author}
  {\bibfnamefont{T.}~\bibnamefont{Terashima}},\ and\ \bibinfo {author}
  {\bibfnamefont{Y.}~\bibnamefont{Matsuda}},\ }%
  \bibfield{journal}{%
  \bibinfo {journal} {Phys. Rev. B}\ }%
  \textbf{\bibinfo {volume} {83}},\ \bibinfo {pages} {060505} (\bibinfo {month}
  {Feb}\ \bibinfo {year} {2011})%
  \bibAnnoteFile{NoStop}{Kasahara-Fermi-liquid-CaFe2As2}%
\bibitem{Furukawa-CaFe2As2}%
  \BibitemOpen
  \bibfield{author}{%
  \bibinfo {author} {\bibfnamefont{Y.}~\bibnamefont{Furukawa}}, \bibinfo
  {author} {\bibfnamefont{B.}~\bibnamefont{Roy}}, \bibinfo {author}
  {\bibfnamefont{S.}~\bibnamefont{Ran}}, \bibinfo {author}
  {\bibfnamefont{S.~L.}\ \bibnamefont{Bud'ko}},\ and\ \bibinfo {author}
  {\bibfnamefont{P.~C.}\ \bibnamefont{Canfield}},\ }%
  \bibfield{journal}{%
  \bibinfo {journal} {Phys. Rev. B}\ }%
  \textbf{\bibinfo {volume} {89}},\ \bibinfo {pages} {121109} (\bibinfo {month}
  {Mar}\ \bibinfo {year} {2014})%
  \bibAnnoteFile{NoStop}{Furukawa-CaFe2As2}%
\bibitem{Pierre-ARPES-review}%
  \BibitemOpen
  \bibfield{author}{%
  \bibinfo {author} {\bibfnamefont{P.}~\bibnamefont{Richard}}, \bibinfo
  {author} {\bibfnamefont{T.}~\bibnamefont{Sato}}, \bibinfo {author}
  {\bibfnamefont{K.}~\bibnamefont{Nakayama}}, \bibinfo {author}
  {\bibfnamefont{T.}~\bibnamefont{Takahashi}},\ and\ \bibinfo {author}
  {\bibfnamefont{H.}~\bibnamefont{Ding}},\ }%
  \bibfield{journal}{%
  \bibinfo {journal} {Rep. Prog. Phys.}\ }%
  \textbf{\bibinfo {volume} {74}},\ \bibinfo {pages} {124512} (\bibinfo {year}
  {2011})%
  \bibAnnoteFile{NoStop}{Pierre-ARPES-review}%
\bibitem{Peng-curvature}%
  \BibitemOpen
  \bibfield{author}{%
  \bibinfo {author} {\bibfnamefont{P.}~\bibnamefont{Zhang}}, \bibinfo {author}
  {\bibfnamefont{P.}~\bibnamefont{Richard}}, \bibinfo {author}
  {\bibfnamefont{T.}~\bibnamefont{Qian}}, \bibinfo {author}
  {\bibfnamefont{Y.-M.}\ \bibnamefont{Xu}}, \bibinfo {author}
  {\bibfnamefont{X.}~\bibnamefont{Dai}},\ and\ \bibinfo {author}
  {\bibfnamefont{H.}~\bibnamefont{Ding}},\ }%
  \bibfield{journal}{%
  \bibinfo {journal} {Rev. Sci. Instrum.}\ }%
  \textbf{\bibinfo {volume} {82}},\ \bibinfo {pages} {043712} (\bibinfo {year}
  {2011})%
  \bibAnnoteFile{NoStop}{Peng-curvature}%
\bibitem{Kaminski-BaFe2As2-CaFe2As2}%
  \BibitemOpen
  \bibfield{author}{%
  \bibinfo {author} {\bibfnamefont{T.}~\bibnamefont{Kondo}}, \bibinfo {author}
  {\bibfnamefont{R.~M.}\ \bibnamefont{Fernandes}}, \bibinfo {author}
  {\bibfnamefont{R.}~\bibnamefont{Khasanov}}, \bibinfo {author}
  {\bibfnamefont{C.}~\bibnamefont{Liu}}, \bibinfo {author}
  {\bibfnamefont{A.~D.}\ \bibnamefont{Palczewski}}, \bibinfo {author}
  {\bibfnamefont{N.}~\bibnamefont{Ni}}, \bibinfo {author}
  {\bibfnamefont{M.}~\bibnamefont{Shi}}, \bibinfo {author}
  {\bibfnamefont{A.}~\bibnamefont{Bostwick}}, \bibinfo {author}
  {\bibfnamefont{E.}~\bibnamefont{Rotenberg}}, \bibinfo {author}
  {\bibfnamefont{J.}~\bibnamefont{Schmalian}}, \bibinfo {author}
  {\bibfnamefont{S.~L.}\ \bibnamefont{Bud'ko}}, \bibinfo {author}
  {\bibfnamefont{P.~C.}\ \bibnamefont{Canfield}},\ and\ \bibinfo {author}
  {\bibfnamefont{A.}~\bibnamefont{Kaminski}},\ }%
  \bibfield{journal}{%
  \bibinfo {journal} {Phys. Rev. B}\ }%
  \textbf{\bibinfo {volume} {81}},\ \bibinfo {pages} {060507} (\bibinfo {month}
  {Feb}\ \bibinfo {year} {2010})%
  \bibAnnoteFile{NoStop}{Kaminski-BaFe2As2-CaFe2As2}%
\bibitem{LDA+DMFT-licht}%
  \BibitemOpen
  \bibfield{author}{%
  \bibinfo {author} {\bibfnamefont{A.~I.}\ \bibnamefont{Lichtenstein}}\ and\
  \bibinfo {author} {\bibfnamefont{M.~I.}\ \bibnamefont{Katsnelson}},\ }%
  \bibfield{journal}{%
  \Doi{10.1103/PhysRevB.57.6884}{\bibinfo {journal} {Phys. Rev. B}}\ }%
  \textbf{\bibinfo {volume} {57}},\ \bibinfo {pages} {6884} (\bibinfo {month}
  {Mar}\ \bibinfo {year} {1998})%
  \bibAnnoteFile{NoStop}{LDA+DMFT-licht}%
\bibitem{LDA+DMFT-anisimov-1997}%
  \BibitemOpen
  \bibfield{author}{%
  \bibinfo {author} {\bibfnamefont{V.~I.}\ \bibnamefont{Anisimov}}, \bibinfo
  {author} {\bibfnamefont{A.}~\bibnamefont{Poteryaev}}, \bibinfo {author}
  {\bibfnamefont{M.}~\bibnamefont{Korotin}}, \bibinfo {author}
  {\bibfnamefont{A.}~\bibnamefont{Anokhin}},\ and\ \bibinfo {author}
  {\bibfnamefont{G.}~\bibnamefont{Kotliar}},\ }%
  \bibfield{journal}{%
  \bibinfo {journal} {J. Phys.: Condens. Matter}\ }%
  \textbf{\bibinfo {volume} {9}},\ \bibinfo {pages} {943} (\bibinfo {year}
  {1997})%
  \bibAnnoteFile{NoStop}{LDA+DMFT-anisimov-1997}%
\bibitem{cRPA-DMFT-LaOFeAs-markus}%
  \BibitemOpen
  \bibfield{author}{%
  \bibinfo {author} {\bibfnamefont{M.}~\bibnamefont{Aichhorn}}, \bibinfo
  {author} {\bibfnamefont{L.}~\bibnamefont{Pourovskii}}, \bibinfo {author}
  {\bibfnamefont{V.}~\bibnamefont{Vildosola}}, \bibinfo {author}
  {\bibfnamefont{M.}~\bibnamefont{Ferrero}}, \bibinfo {author}
  {\bibfnamefont{O.}~\bibnamefont{Parcollet}}, \bibinfo {author}
  {\bibfnamefont{T.}~\bibnamefont{Miyake}}, \bibinfo {author}
  {\bibfnamefont{A.}~\bibnamefont{Georges}},\ and\ \bibinfo {author}
  {\bibfnamefont{S.}~\bibnamefont{Biermann}},\ }%
  \bibfield{journal}{%
  \bibinfo {journal} {Phys. Rev. B}\ }%
  \textbf{\bibinfo {volume} {80}},\ \bibinfo {pages} {085101} (\bibinfo {month}
  {Aug}\ \bibinfo {year} {2009})%
  \bibAnnoteFile{NoStop}{cRPA-DMFT-LaOFeAs-markus}%
\bibitem{cRPA-ferdi-2004}%
  \BibitemOpen
  \bibfield{author}{%
  \bibinfo {author} {\bibfnamefont{F.}~\bibnamefont{Aryasetiawan}}, \bibinfo
  {author} {\bibfnamefont{M.}~\bibnamefont{Imada}}, \bibinfo {author}
  {\bibfnamefont{A.}~\bibnamefont{Georges}}, \bibinfo {author}
  {\bibfnamefont{G.}~\bibnamefont{Kotliar}}, \bibinfo {author}
  {\bibfnamefont{S.}~\bibnamefont{Biermann}},\ and\ \bibinfo {author}
  {\bibfnamefont{A.~I.}\ \bibnamefont{Lichtenstein}},\ }%
  \bibfield{journal}{%
  \Doi{10.1103/PhysRevB.70.195104}{\bibinfo {journal} {Phys. Rev. B}}\ }%
  \textbf{\bibinfo {volume} {70}},\ \bibinfo {pages} {195104} (\bibinfo {month}
  {Nov}\ \bibinfo {year} {2004})%
  \bibAnnoteFile{NoStop}{cRPA-ferdi-2004}%
\bibitem{TMO-vaugier}%
  \BibitemOpen
  \bibfield{author}{%
  \bibinfo {author} {\bibfnamefont{L.}~\bibnamefont{Vaugier}}, \bibinfo
  {author} {\bibfnamefont{H.}~\bibnamefont{Jiang}},\ and\ \bibinfo {author}
  {\bibfnamefont{S.}~\bibnamefont{Biermann}},\ }%
  \bibfield{journal}{%
  \bibinfo {journal} {Phys. Rev. B}\ }%
  \textbf{\bibinfo {volume} {86}},\ \bibinfo {pages} {165105} (\bibinfo {year}
  {2012})%
  \bibAnnoteFile{NoStop}{TMO-vaugier}%
\bibitem{udyn-werner}%
  \BibitemOpen
  \bibfield{author}{%
  \bibinfo {author} {\bibfnamefont{P.}~\bibnamefont{Werner}}, \bibinfo {author}
  {\bibfnamefont{M.}~\bibnamefont{Casula}}, \bibinfo {author}
  {\bibfnamefont{T.}~\bibnamefont{Miyake}}, \bibinfo {author}
  {\bibfnamefont{F.}~\bibnamefont{Aryasetiawan}}, \bibinfo {author}
  {\bibfnamefont{A.~J.}\ \bibnamefont{Millis}},\ and\ \bibinfo {author}
  {\bibfnamefont{S.}~\bibnamefont{Biermann}},\ }%
  \bibfield{journal}{%
  \Doi{doi:10.1038/nphys2250}{\bibinfo {journal} {Nature Physics}}\ }%
  \textbf{\bibinfo {volume} {8}},\ \bibinfo {pages} {331} (\bibinfo {year}
  {2012})%
  \bibAnnoteFile{NoStop}{udyn-werner}%
\bibitem{Ambroise-BaCo2As2}%
  \BibitemOpen
  \bibfield{author}{%
  \bibinfo {author} {\bibfnamefont{A.}~\bibnamefont{van Roekeghem}}, \bibinfo
  {author} {\bibfnamefont{T.}~\bibnamefont{Ayral}}, \bibinfo {author}
  {\bibfnamefont{J.~M.}\ \bibnamefont{Tomczak}}, \bibinfo {author}
  {\bibfnamefont{M.}~\bibnamefont{Casula}}, \bibinfo {author}
  {\bibfnamefont{N.}~\bibnamefont{Xu}}, \bibinfo {author}
  {\bibfnamefont{H.}~\bibnamefont{Ding}}, \bibinfo {author}
  {\bibfnamefont{M.}~\bibnamefont{Ferrero}}, \bibinfo {author}
  {\bibfnamefont{O.}~\bibnamefont{Parcollet}}, \bibinfo {author}
  {\bibfnamefont{H.}~\bibnamefont{Jiang}},\ and\ \bibinfo {author}
  {\bibfnamefont{S.}~\bibnamefont{Biermann}},\ }%
  \bibfield{journal}{%
  \bibinfo {journal} {Phys. Rev. Lett.}\ }%
  \textbf{\bibinfo {volume} {113}},\ \bibinfo {pages} {266403} (\bibinfo {year}
  {2014})%
  \bibAnnoteFile{NoStop}{Ambroise-BaCo2As2}%
\bibitem{Ambroise-SrVO3}%
  \BibitemOpen
  \bibfield{author}{%
  \bibinfo {author} {\bibfnamefont{A.}~\bibnamefont{van Roekeghem}}\ and\
  \bibinfo {author} {\bibfnamefont{S.}~\bibnamefont{Biermann}},\ }%
  \bibfield{journal}{%
  \bibinfo {journal} {EPL}\ }%
  \textbf{\bibinfo {volume} {108}},\ \bibinfo {pages} {75003} (\bibinfo {year}
  {2014})%
  \bibAnnoteFile{NoStop}{Ambroise-SrVO3}%
\bibitem{udyn-michele}%
  \BibitemOpen
  \bibfield{author}{%
  \bibinfo {author} {\bibfnamefont{M.}~\bibnamefont{Casula}}, \bibinfo {author}
  {\bibfnamefont{A.}~\bibnamefont{Rubtsov}},\ and\ \bibinfo {author}
  {\bibfnamefont{S.}~\bibnamefont{Biermann}},\ }%
  \bibfield{journal}{%
  \Doi{10.1103/PhysRevB.85.035115}{\bibinfo {journal} {Phys. Rev. B}}\ }%
  \textbf{\bibinfo {volume} {85}},\ \bibinfo {pages} {035115} (\bibinfo {month}
  {Jan}\ \bibinfo {year} {2012})%
  \bibAnnoteFile{NoStop}{udyn-michele}%
\bibitem{SrVO3-dynU-Wang}%
  \BibitemOpen
  \bibfield{author}{%
  \bibinfo {author} {\bibfnamefont{L.}~\bibnamefont{Huang}}\ and\ \bibinfo
  {author} {\bibfnamefont{Y.}~\bibnamefont{Wang}},\ }%
  \bibfield{journal}{%
  \bibinfo {journal} {EPL}\ }%
  \textbf{\bibinfo {volume} {99}},\ \bibinfo {pages} {67003} (\bibinfo {year}
  {2012})%
  \bibAnnoteFile{NoStop}{SrVO3-dynU-Wang}%
\bibitem{Shell-folding}%
  \BibitemOpen
  \bibfield{author}{%
  \bibinfo {author} {\bibfnamefont{P.}~\bibnamefont{Seth}}, \bibinfo {author}
  {\bibfnamefont{P.}~\bibnamefont{Hansmann}}, \bibinfo {author}
  {\bibfnamefont{A.}~\bibnamefont{van Roekeghem}}, \bibinfo {author}
  {\bibfnamefont{L.}~\bibnamefont{Vaugier}},\ and\ \bibinfo {author}
  {\bibfnamefont{S.}~\bibnamefont{Biermann}},\ }%
  \bibfield{journal}{%
  \bibinfo {journal} {in preparation }}%
   (\bibinfo {year} {2015})%
  \bibAnnoteFile{NoStop}{Shell-folding}%
\bibitem{TRIQS-website}%
  \BibitemOpen
  \bibfield{author}{%
  \bibinfo {author} {\bibfnamefont{M.}~\bibnamefont{Ferrero}}\ and\ \bibinfo
  {author} {\bibfnamefont{O.}~\bibnamefont{Parcollet}},\ }%
  \enquote{\bibinfo {title} {{TRIQS}: A toolbox for research on interacting
  quantum systems},}\  (\bibinfo {year} {2011}),\
  \url{http://ipht.cea.fr/triqs}%
  \bibAnnoteFile{NoStop}{TRIQS-website}%
\end{thebibliography}%

\end{document}